\documentclass[aps,twocolumn,showpacs,preprintnumbers,amsmath,amssymb]{revtex4-1}
\usepackage{array}
\newcolumntype{L}[1]{>{\raggedright\let\newline\\\arraybackslash\hspace{0pt}}m{
#1}}
\newcolumntype{C}[1]{>{\centering\let\newline\\\arraybackslash\hspace{0pt}}m{#1}
}
\newcolumntype{R}[1]{>{\raggedleft\let\newline\\\arraybackslash\hspace{0pt}}m{#1
}}

\usepackage{mathrsfs}
\usepackage[dvips]{graphicx}
\usepackage{mathtools}
\usepackage{float}
\usepackage{bm}
\usepackage[usenames, dvipsnames]{color}
\usepackage{amsmath}
\definecolor{orange}{RGB}{255,120,0}
\definecolor{gold}{RGB}{130,100,0}
\definecolor{maroon}{RGB}{128,0,50}
\definecolor{red}{RGB}{255,0,0}
\numberwithin{equation}{section}

\newcommand{\mb}{\mathbf}

\newcommand{\m}{\mathrm}
\newcommand{\h}{\hbar}

\newcommand{\p}{\mathbf{k}}
\newcommand{\ee}{\varepsilon}
\newcommand{\kk}{\mathbf{k}}
\newcommand{\q}{\mathbf{q}}

\newcommand{\la}{\langle}
\newcommand{\ra}{\rangle}
\newcommand{\cd}{\cdot}
\newcommand{\ph}{\phantom\dag}
\newcommand{\rr}{\mathbf{r}}
\newcommand{\RR}{\mathbf{R}}

\newcommand{\e}{\varepsilon}
\newcommand{\s}{\sigma}
\newcommand{\ld}{\lambda}
\newcommand{\om}{\omega}
\newcommand{\al}{\alpha}
\newcommand{\D}{\Delta}

\def\mathclap#1{\text{\hbox to 0pt{\hss$\mathsurround=0pt#1$\hss}}}
\begin{document}
\def \brho{{\hbox{\boldmath $\rho$}}}
\def \beps{{\hbox{\boldmath $\epsilon$}}}
\def \bdelta{{\hbox{\boldmath $\delta$}}}
\title{Semianalytical study of excitons and quasiparticle band gap in two-dimensional insulators}
\author{Zoran Rukelj$^{1}$}
\email{zrukelj@phy.hr}
\author{Vito Despoja$^{2}$}
\affiliation{$^1$Department of Physics, University of Zagreb, Bijeni\v{c}ka 32, HR-10000 Zagreb, Croatia}
\affiliation{$^2$ Institute of Physics,  Bijeni\v{c}ka 46, HR-10000 Zagreb, Croatia}

\begin{abstract}
\noindent

{
A theoretical study of the exciton binding energy in the two-dimensional hexagonal boron nitride monolayer is presented within the tight-binding approximation (TBA).
A self-consistent equation for the interband electron-hole propagators is derived and in the long wavelength limit reduced to the standard hydrogen atom like Schr\"{o}dinger equation. 
It is shown that inclusion of dynamically screened Coulomb interaction in ladder term is of crucial importance for proper description of exciton binding energy. 
This leads to the self-consistent eigenvalue problem with dynamical screening. 
The dependence of the exciton energy on the orbital quantum number is studied. It is predicted 
that for the fixed principal quantum number the states with higher orbital momentum have lower energy than the
states with lower orbital momentum. Using the developed formulas and the experimental optical gap 
the quasiparticle gap is estimated. In the limit of high polarizability, a semiclassical procedure was used to obtain the  exciton  binding energy.
The TBA parametrization is supported by {\em ab initio} calculations. }

\end{abstract}
\maketitle

\section{Introduction}

For the last decade, the condensed matter physics has been dominated by experimental and theoretical investigation of the two-dimensional materials.
One  class of these materials are the direct gap two-dimensional insulators. The best known 
examples are monolayers (ML) of the members of transition-metal dichalcogenides  \cite{hbn,mos3d} and a  hexagonal boron nitride  monolayer (hBN-ML).
hBN-ML, the  simplest of the 2D insulators, exhibits  a wide optical 
gap which in turn is responsible for a low electronic polarizability. With this 
electronic property comes the chemical inertness and mechanical stability \cite{lee,mehanika}, 
which originates from the strong  $\s$ bonds between the $sp^2$ hybrids.  
However, in spite of its  simple structure, some  basic problems still remain unsolved. 
The most significant is the inability to experimentally determine the single particle band gap, 
which numerous {\it ab initio} studies estimate to be between $7$ and $9$ eV \cite{guo1,wirt}.
On the other hand, well established {\it ab initio} methodology which includes quasiparticle corrections (GW method) and 
solving the Bethe-Salpeter equation (BSE), so called GW-BSE method, is capable of giving accurate single particle and optical gap energies, including excitonic energy and their
oscillator strength \cite{Thygesen,LouieMoS2, koskelo}. However, the GW-BSE method is still computationally very heavy and time consuming (even for 2D crystals) 
while the method here proposed is semi-analytical and provides a fast  estimation of excitonic  energy and single particle gap.

In three-dimensionsional insulators, the  exciton energies  can be calculated using the simple Wannier hydrogen-like  
model \cite{wannier} in which the Coulomb interaction is screened by dielectric constant. In the optical absorption measurements the lowest exciton state appears as a 
well-defined isolated peak and higher exciton energies rapidly converge toward the 
single particle band gap \cite{cuo2}. Therefore, the error in estimating the  single particle band gap is relatively small.

However, in two-dimensional insulators the screening effects are large due to  spatially depended  dielectric function which screens the Coulomb potential in a nontrivial way.
This modifies the hydrogen-like model substantially \cite{chernikov}, resulting in (compared with the 3D case) a drastic change of 
quantized exciton energies and the absence of the degeneracy of the hydrogen-like orbitals, leading to their dependence not only on the 
principal quantum number $n$, but on the orbital quantum number $\ell$ as well.
The $\ell$-dependence of the exciton energies can
be seen  by probing the two-dimensional insulators with the two photon luminescence techniques \cite{exi-naj}.

In this paper the single particle properties of the hBN-ML  are calculated using the tight binding approximation.
The TBA parameters are determined using the conductivity sum rule and from the difference of the
bare atomic orbitals energies. The calculated TBA parameters are supported by the {\it ab initio} results such as DFT-LDA band gap and  effective masses.
The electronic polarizabilites have been calculated in both approaches and compared. Also within the TBA model the charge vertices have been derived explicitly. This will allows us to classify various contributions to the election-hole propagator equation. 
The exciton energies are derived using the equation of motion techniques for the electron-hole pair propagator \cite{kohn, wiser}.  
A systematic derivation of the four contributions (in the leading order of Coulomb interaction)  to the election-hole propagator equation have been analyzed. In the ladder part,
which governs the electron-hole dynamics, the dynamically screened Coulomb interaction is implemented. In the long wave limit approximation the electron-hole equation is reduced to Schr\"{o}dinger equation
for dynamical screened potential  which is then solved self-consistently. The eigenvalues, i.e. the exciton energies, are calculated for the dynamical and statical case and compared.
It will be argued that it is possible to obtain a realistic value of the single particle band gap using the experimentally obtained energy of the exciton ground state.  
This can be done by combining the DFT-LDA  calculations of the static polarizability and the effective mass of the electron in the valence ($v$) and the conduction ($c$) band around the 
K point, which enter in to  the Schr\"{o}dinger equation with the  screened Coulomb potential.
Here it is assumed that hBN-ML sheet is free-standing.  This way, various effects originating from the finite substrate polarization are deliberately avoided \cite{band-reno}.
Particularly, the exciton ground state energy in the high polarizability limit 
is investigated by employing the Einstein-Brillouin-Keller (EBK) procedure \cite{ebk1} and compared with the solutions of the Schr\"{o}dinger equation for the screened Coulomb potential.
This leads to the analytical expression for the exciton ground state energy which is used to predict the quasiparticle band gaps of the selected two-dimensional 
insulators\cite{gapovi1,gapovi2,gapovi3}.

The extended calculations of the quasiparticle  and exciton properties in hBN-ML followed by the {\it ab initio} parametrized TBA is given in  
Ref.\cite{jeb}. But unlike the results presented in this paper, their results are to extensive but not applicable for a simple estimation of the exciton energies and the quasiparticle 
band gap.

This paper is organized as follows.
In Sec. \ref{band} the electronic band structure is determined using {\em ab initio}  and tight binding approximation. A
brief formulation of the {\em ab initio} calculation of the dielectric response and quasiparticle corrections in wide gap 2D crystals is presented. 
The band gap is estimated within the same {\it ab initio} formalism and using the conductivity sum rule.
In Sec. \ref{jedbe}, the equations that determine the dynamics of the electron-hole propagation are derived to the first order in Coulomb interaction.
This equation is equivalent to the two-body Schr\"{o}dinger equation and is  solved for the case of bare and dynamically screened Coulomb interaction. 
The screened Coulomb interaction is calculated within the two-band TBA model.
In Sec.\ref{rez} the results are presented. The first ten  exciton energy levels and their spatial extend are calculated.
The exciton ground state energy in the high polarizability limit 
is investigated by using the EBK procedure  and compared with the results of Sec.\ref{rez}.

\section{Band structure}\label{band}
\subsection{{\it Ab initio} studies of hBN-ML band structure and dielectric properties}\label{dft-vito}
\label{abin}
In order to maintain the TBA parametrization in the framework of 
realistic crystal values the DFT calculation of the electronic ground 
state and the RPA dielectric function of hBN-ML are provided. Additionally, a brief quasiparticle 
G$_0$W$_0$ correction of  DFT-LDA band gap around K point is provided, as discussed later.         

At the DFT stage of the calculation the Kohn-Sham (KS) wave functions $\varphi_{L\kk}(\rr)$ and energy 
levels $E_{\kk}^L$, i.e. the band structure of a hBN-ML is determined using the 
plane-wave DFT code  {\sc Quantum ESPRESSO} (QE)~\cite{QE}. 
The core-electron interaction is approximated by the norm-conserving pseudopotentials~\cite{pseudopotentials},  and the exchange correlation (XC) potential by 
the LDA Perdew-Zunger (PZ) functional~\cite{PZ}. 
For the hBN-ML primitive cell constant, $a = 4.746 \,a_0$ ($a_0$ is the Bohr radius) is used and the superlattice constant in the $z$ direction is $L=23.73 \, a_0$. 
The ground state electronic densities of the hBN-ML are calculated 
using the $12\times 12\times1$ Monkhorst-Pack {\bf k}-point mesh~\cite{MPmesh} of the first Brillouin zone (BZ). 
For the plane-wave cut-off energy $60$~Ry ($816$~eV). In order to obtain sharp Van Hove 
singularities the partial density of states (PDOS) are calculated using the $101\times 101\times1$ Monkhorst-Pack {\bf k}-point mesh.

If hBN-ML is approximated as fully 2D system its dielectric function is given by (\ref{sed2})
with the independent electrons response function $\chi^{0}(\q,\omega)=L\chi_{g_z=0,g_z'=0}^{0}(\q,\omega)$ given by (\ref{sed3}).
However, if the dispersivity of the dielectric response in the direction 
perpendicular to crystal lattice plane (the $z$ direction) is included, the response function matrix becomes 
\begin{eqnarray}
\chi_{g_z,g_z'}^{0}(\q,\omega)=\hspace{6cm}
\label{Resfun0}\\
\frac{1}{V}\sum_{\kk \s LL'} G^{LL'}_{\kk,\kk+\q}(g_z) \frac{f_{\kk}^L-f_{\kk+\q}^{L'}}
{\hbar\omega+i\eta+E_{\bf k}^L-E_{{\bf k}+{\bf q}}^{L'}}
{G^*}^{LL'}_{\kk,\kk+\q}(g_z'),
\nonumber
\end{eqnarray}  
where $f^L_{{\kk}}$ is the Fermi-Dirac distribution at 
temperature $T$ and the charge vertices are 
\begin{equation}
G^{LL'}_{\kk,\kk+\q}(g_z)= \int_V d^3\rr \varphi^*_{L{\kk}}(\rr)
e^{-i \q \cd \boldsymbol{\rho}}e^{-ig_zz}\varphi_{L'{\bf k}+{\bf q}}(\rr).
\label{Matrel}
\end{equation}
Here ${\bf q}$ is the momentum transfer vector parallel to the $x-y$ plane 
and ${\bf r}=(\brho,z)$ is a $3D$ position vector and $g_z$ is the  reciprocal lattice 
vector in the perpendicular ($z$) direction. Integration in \eqref{Matrel} is performed over the normalization volume $V=S\times L$, where $S$ is the normalization 
surface. The independent electron response function \eqref{Resfun0} is calculated 
using $201\times 201\times1$ $\bf k$-point mesh sampling which corresponds to $40405$ Monkhorst-Pack special $\bf k$-points in the Brillouin zone. This $\bf k$-point mesh sampling enables the minimum transfer wave 
vector $q_{min}=0.0076$~$a_0^{-1}$. 
The damping parameter used is 
$\eta=50$~meV and the temperature is $k_BT=10$~meV. The band summation is performed over $40$ bands, which proved to be sufficient for proper description of the electronic excitations up to $30$~eV. 

It is shown that in the long wavelength limit ($\q \approx 0$) the hBN-ML 2D static dielectric 
function can be approximated as: 
\begin{equation}
\epsilon(\q \approx 0,\omega=0)=1+\lambda_{DFT} |\q|
\label{eps0}
\end{equation}  
where the DFT screening length is $\lambda_{DFT}=10.5 \, a_0$. Considering wide hBN-ML band gap, the static approximation (\ref{eps0}) 
is valid in the dynamical limit, even up to $\hbar \omega=3\,$eV, which is especially useful in further 
estimation of the quasiparticle band gap.  

Because the LDA always underestimates the semiconducting band gap, it is of crucial importance to provide quasiparticle corrections of the band structure in order to obtain the accurate exciton energy.       
Here a brief estimation of the band gap for wide band gap semiconducting layers is proposed. 
Because of the wide  hBN-ML band gap, the dynamically screened Coulomb 
interaction will be approximated by its statical limit 
${{w}_{\q}}(\omega)\approx {{ w}_{\q}}(\omega=0)$. On the other hand, even if the crystal is atomically   
thick, the dispersivity of the statical response in the $z$ direction (inclusion of $g_z, g_z'\ne 0$ 
in (\ref{Resfun0})) plays an important role for the accurate quasiparticle correction.      

The quasiparticle corrections of LDA energies $E^L_{{\bf k}}$ are provided within the Statically 
Screened Exchange Coulomb hole Correlation GW aproximation, usually  called the COH-SEX approximation \cite{Hedin,Louie}
\begin{equation}
\tilde{E}^L_{{\bf k}}=E^L_{{\bf k}}-E^{XC}_{L{\bf k}}+\Sigma^{COH}_{L{\bf k}}+\Sigma^{SEX}_{L{\bf k}}.
\label{Ecorr}
\end{equation}
Here XC is LDA exchange correlation energy, COH correlation energy is
\begin{eqnarray}\label{COH}
\Sigma^{COH}_{L{\bf k}}=\frac{1}{2}\sum_{L' g_zg_z'}\int\frac{d{\bf q}}{(2\pi)^2}
{{w}}_{g_zg_z'}^{ind}
({\bf q},\omega=0)G^{LL'}_{{\bf k},{\bf k}+{\bf q}}(g_z-g_z') \nonumber \\
\end{eqnarray}
and static SEX energy is
\begin{eqnarray}\label{SEX}
\Sigma^{SEX}_{L{\bf k}}=
-\sum_{L' g_zg_z'}\int\frac{d{\bf q}}{(2\pi)^2}f^{L'}_{{\bf k}+{\bf q}}\times\hspace{3cm}
\\
{{w}}_{g_z,g_z'}({\bf q},\omega=0)
{G}^{LL'}_{\kk,\kk+\q}(g_z){G^*}^{LL'}_{\kk,\kk+\q}(g_z').
\nonumber
\end{eqnarray}
The induced Coulomb interaction matrix $\hat{w}^{ind}=\hat{v}\hat{\chi}\hat{v}$
is determined by solving Dyson-matrix equation $\hat{\chi}=\hat{\chi}^0+
\hat{\chi}^0\hat{v}\hat{\chi}$ for the screened 
response matrix $\hat{\chi}$. The bare Coulomb interaction matrix elements are given by
\[
v_{g_{z} g_z'}({\bf q})=\frac{v_{\q}}{L}\int^{L/2}_{-L/2}dzdz'
e^{-q|z-z'|}e^{ig_zz}e^{-ig_z'z'}
\]
and total, screened Coulomb interaction matrix is $\hat{w}=\hat{v}+\hat{{w}}^{ind}$.

The hBN-ML is a direct gap insulator with the conduction band minimum (CBM)  and the valence band 
maximum (VBM) located at the K point of the Brillouin zone. The DFT band gap obtained in this calculation is $2\Delta_{DFT}=6\,$eV and after the quasiparticle 
correction (\ref{Ecorr}--\ref{SEX})  (which is for this purpose provided just in K point of the Brillouin zone)  it increases to $2\Delta_{GW}=9\,$eV.

The hBN-ML band structure and PDOS are shown in Fig.\ref{f1}(b) with an emphasis on
the conduction and the valence band. The primary and secondary minima in the conduction 
band are only $0.115$ eV apart. In the 3D case the Van der Waals interaction shifts this second minimum below the first one, thus making  
hBN an indirect gap insulator \cite{indirekt}. 

In order to better understand the electronic properties, the orbital decomposition 
of the valence bands through the PDOS calculation is also provided, shown in 
Fig.\ref{f1}(b). It can be seen that in the vicinity of the K point, the valence band is formed 
entirely from the boron $2p_z$ orbital while the conduction
band is formed entirely from the nitrogen $2p_z$ orbital. 
The calculated effective masses of the conduction and the valence band at K point are 
$m^*_c = 0.8m_e$ and $m^*_v = 0.75m_e$, respectively. The effective mass of the 
valence band at the ${\Gamma}$ point is found to be approximately equal to the electron mass $m_e$.

\subsection{The tight binding approximation }\label{TBA}

 \begin{figure*}[tt]
\includegraphics[width=6.cm]{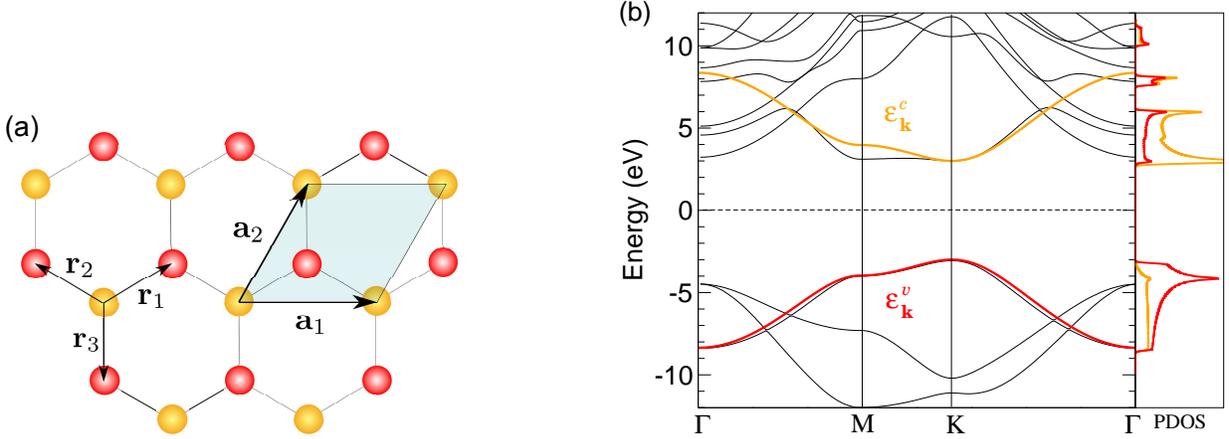} 
\hspace{1.9cm}
\includegraphics[width=8.1cm]{hbn.eps}
\caption{\small{(a) The crystal structure of hBN-ML with the nitrogen (red) and the boron (yellow) atoms in the base spanned by primitive vectors $\mb{a}_1$ and $\mb{a}_2$. (b) 
The band structure of hBN-ML obtained by the {\it ab initio} calculation together with the PDOS showing the 
contributions from the nitrogen $2p_z$ orbital (red) and the boron $2p_z$ orbital (yellow). The TBA band structure of the valence (red) and conduction (orange) 
bands (\ref{e-bloch})}.}
\label{f1}
\end{figure*}

The band structure of hBN-ML can be easily  obtained using the TBA model, which is  here presented in the second quantization representation. 
The fermionic field operator in the basis of localized atomic orbitals labeled by index $n$ is 
\begin{equation} \label{psi}
 \Psi^{\dag}(\rr) = \sum_{ n \s \RR}c^{\dag}_{n \s \RR}    \phi^*_{n \s}(\mb{r} - \mb{R}).
\end{equation}
In the case of hBN-ML the two $2p_z$ orbitals on nitrogen and boron
atoms are labeled as $n  = N$ or $B$, respectively with the spin index $\s$.
Vectors $\mb{R}$ describe an absolute position of the atomic orbital  $n$ in the crystal and are usually decomposed as  $\mb{R} = \RR_0 + \mb{r}_n $,
with  $\RR_0$ being the primitive lattice vector and the 
$\mb{r}_n$ the position of the orbital $n$ within the primitive cell.
It is assumed that the atomic orbitals are localized such that 
\begin{equation} \label{lokalizacija}
\int d\mb{r} \phi^*_{n\s}(\mb{r} - \mb{R})\phi_{n'\s'}(\mb{r} - \mb{R'})  =  \delta_{n n' } \delta_{\s \s' } \delta_{\RR \RR' } .
\end{equation}
The bare  Hamiltonian operator is  defined as
\begin{equation} \label{ham1}
\hat{H}_0 = \int d\mb{r} \Psi^{\dag}(\rr) \hat{H}(\rr) \Psi(\rr),
\end{equation}
where the real space  Hamiltonian $\hat{H}(\rr)$
consists of the bare atomic Hamiltonian and the contribution  from the 
residual two-particle interaction. The latter is described by an effective single particle interaction called the crystal potential 
\cite{barisic} or the effective potential, from the DFT point of view. 
Implementing  (\ref{psi}) and (\ref{lokalizacija}) in (\ref{ham1}), yields
\begin{equation} \label{ham2}
\hat{H}_0 = \sum_{n  n' \s}\sum_{\RR j} H^{nn'}(\rr_j)c^{\dag}_{n \RR \s}c^{\ph}_{n' \RR+\rr_j \s}. 
\end{equation}
Since the matrix elements of the Hamiltonian  (\ref{ham2})  depend  on the distance between atomic orbitals,
a set of vectors $\{ \rr_j \}$ are introduced which  represent the relative positions of the first neighbors, with the zero vector ($\rr=0$) formally included, as shown in Fig.\ref{f1}(a).
Thus the  diagonal elements in (\ref{ham2})  are the  atomic orbital energies $\e_{n}$,  $H^{NN}(0) = \ee_N$ , $H^{BB}(0) = \ee_B$ and
the off-diagonal elements $H^{NB}(\rr_j) = H^{BN}(\rr_j)=-t_0$, give the electron hopping parameter $t_0$. 
Next, the Hamiltonian (\ref{ham2}) is written in  the delocalized orbitals representation $\{ n \kk\}$ by transforming  the fermion operators
\begin{equation}\label{trans1}
 c^{\dag}_{n \kk \s}= \frac{1}{\sqrt{N}}\sum_{\RR}{\rm{e}}^{i\kk \cd \RR} c^{\dag}_{n \RR \s}.
\end{equation}
Thus, the Hamiltonian transforms as 
\begin{equation}\label{ham3}
 \hat{H}_0 = \sum_{nn'}\sum_{\kk \s} H^{nn'}_{\kk}c^{\dag}_{n \kk \s}c^{\ph}_{n' \kk \s}, 
\end{equation}
where the  matrix elements are 
\begin{eqnarray} \label{elementi}
&&\hspace{0mm} H^{NN}_{\kk}=\ee_N, \hspace{4mm}  H^{NB}_{\kk}= t_{\kk} \nonumber \\
&&\hspace{0mm}  H^{BN}_{\kk}=t^*_{\kk}, \hspace{4mm}  H^{BB}_{\kk}=\ee_B,
\end{eqnarray}
and the term $ t_{\kk}$ is defined as
\begin{equation}\label{tk}
 t_{\kk} = -t_0\sum_{j=1,2,3}\m{e}^{-i \kk \cd \rr_j}.
\end{equation} 

The transition from the delocalized orbital representation $\{n \kk \}$ to the diagonal Bloch representation $\{L \kk \}$ is obtained by a unitary transformation 
\begin{equation} \label{trans2}
 c^{\dag}_{L \kk \s} =  \sum_{n}   U_{\kk}(L,n)c^{\dag}_{ n \kk \s}. 
\end{equation}
The inverse transformation of (\ref{trans2}) can be defined as
\begin{equation} \label{trans3}
c^{\dag}_{n \kk \s}=  \sum_{L}   V_{\kk}(n,L) c^{\dag}_{L \kk \s},
\end{equation}
where $\mb{V}=\mb{U}^{-1}$ is unitary matrix inverse.
The elements of matrices $\mb{U}$ and $\mb{V}$ are presented in Appendix \ref{appA}.
The diagonalization of (\ref{ham3}) is straightforward. Introducing  $2\Delta = \varepsilon_{B} - \varepsilon_{N}$
and choosing  $\varepsilon_{B} + \varepsilon_{N} = 0$, gives
\begin{equation} \label{e-bloch}
\ee_{\kk}^{c,v} = \pm  \sqrt{ \Delta^2 + |t_{\kk}|^2  }. 
\end{equation}
It can be noted that in the spacial case of two identical atoms in the base,  $\Delta = 0$, and the eigenvalues (\ref{e-bloch}) reduce to graphene 
dispersions \cite{walace}
\begin{equation} \label{te}
 \ee_{\kk}^{c,v} = \pm |t_{\kk}| = \pm t_0\sqrt{ 3 + 2\m{cos} k_xa + 4\m{cos}\frac{k_xa}{2}\m{cos}\frac{\sqrt{3}k_ya}{2}   }.
\end{equation}
After the diagonalization, the Hamiltonian  (\ref{ham3}) has the simple form
\begin{equation} \label{dijag}
 \hat{H}_0 = \sum_{L \kk \s} \e_{\kk}^Lc^{\dag}_{L \kk \s}c^{\ph}_{L \kk \s}.
\end{equation}

\subsection{Determination of the TBA band parameters}\label{parametri}

The electron hopping parameter $t_0$ can be determined by fitting the dispersions (\ref{e-bloch}) to the {\it ab initio} results  shown in  Fig.\ref{f1}(b),
or using the conductivity sum rule \cite{mahan,nozier}. This rule is applicable to all cases in which the electron 
states at the bottom or at the top of the band ($\Gamma$ point of the Brillouin zone in our case) have the quadratic dispersion and do not exhibit hybridization 
with the states from other bands. 
This is clearly the case for the valence  band, since the low-lying bands 
are formed from the $\s$ bonds between the planar $sp^2$ hybrids. Hence, any matrix element containing transition between the planar hybrid 
and the $2p_z$ orbitals vanishes due to symmetry.

The conductivity sum rule states that the total spectral weight of the total conductivity tensor with Cartesian 
indices $\al$ can be connected with the plasmon dispersion relation $\Omega_{pl}(\q)$ as
\begin{equation}
 4\int_{-\infty}^{\infty} \Re \s_{\al \al}^{tot}(\om,\q)\, d\om = \Omega_{pl}^2(\q) \approx \frac{2\pi n^{tot}_{\al \al} |\q|}{m_e}. 
\end{equation}
The $n^{tot}_{\al \al} = (1/V)\sum_{L \kk \s}f^L_{\kk}$ is the total concentration of the conducting electrons and  can be decomposed as $n^{tot}_{\al \al} = n^{inter}_{\al \al} + n_{\al \al}^{intra}$.
The effective intraband concentration of conducting electrons \cite{kup} is defined as $n_{\al \al}^{intra} = (1/V) (m_e/\hbar^2)\sum_{L\kk \s} ( \partial^2 \ee_{\kk}^L/ \partial^2 k_{\al}) f_{\kk}^L $.
For vanishing low electron concentration in the  valence  band, by definition  $n_{\al \al}^{intra} \leq n^{tot}_{\al \al}$ and  from their explicit forms follows
\begin{equation} \label{1m}
  \frac{1}{m^*_{\al \al}}  = \frac{1}{\hbar^2} \frac{\partial^2 \ee_{\kk}^v}{\partial k_{\al}^2 } \bigg |_{\Gamma}  \leq \frac{1}{m_e}.
\end{equation}
The effective mass tensor is diagonal and isotropic for the TBA dispersions, i.e. $m^*_{\al \al} = m^*$, where  
\begin{equation} \label{effe}
\frac{1}{m^*} = \frac{3}{2\hbar^2}\frac{t_0^2a^2}{\sqrt{\D^2 + 9t_0^2}}.
\end{equation}
From the assumption that the orbital energies of boron and nitrogen are equal to the bare atomic ones  ($\ee_n \approx \ee^0_n$), follows 
$2\D \approx 6\,$eV \cite{atomske-tablice}.
If the lower limit of (\ref{1m}) is taken, i.e.  $m^*\approx m_e$, with  the lattice parameter $a = 4.746 \,a_0$, the expression (\ref{effe}) gives  $t_0 \approx 2.6$ eV. 

The TBA bands (\ref{e-bloch}) with the derived parameters $t_0$ and $\Delta$ are shown in the Fig.\ref{f1}(b). The agreement between the TBA bands and the ones 
obtained from many-body DFT calculations (which includes Hartree-Fock contribution)  may come as a surprise.
Here, these many particle correlation effects have been phenomenologically incorporated in the simple single particle TBA model by choosing the adequate value of the
atomic orbital energies $\ee_n$ and the lower limit of (\ref{1m}). 

The TBA effective masses of the valence and conductive bands at the K point 
(which will be used in the following calculations) are
\begin{equation} \label{effe2}
m^*_c=m^*_v = 4 \Delta \hbar^2/(3t_0^2a^2),
\end{equation}
or $m^*_c=m^*_v = 0.75 m_e$.

\subsection{The charge density operator}
The charge density operator 
\begin{equation}\label{ro}
\hat{\varrho}(\rr)= e\Psi^{\dag}(\rr)\Psi(\rr)
\end{equation}
can be derived within the simple two-band TBA model presented in the section \ref{TBA}.
Using condition (\ref{lokalizacija}) and Fourier transformations (\ref{ro}) the charge density operator in 
$\q \approx 0$ becomes
\begin{eqnarray} \label{ro2}
 \hat{\varrho}(\q) &=&  \sum_{\RR \RR'} \sum_{n n'} \sum_{\s \s'} c^{\dag}_{n \s \RR} c^{\ph}_{n'\s' \RR'}  \nonumber \\
 && \times  e \int  d \rr \phi_{n\s}^*(\rr-\RR) {\rm{e}}^{-i\q \cd \rr} \phi_{n'\s'}(\rr-\RR') \nonumber \\
                   &\approx&  \sum_{\RR} {\rm{e}}^{-i\q \cd \RR} \sum_{\s} \sum_{n n'}e \delta_{n n'} c^{\dag}_{n \s \RR} c^{\ph}_{n' \s \RR}.
\end{eqnarray}
With an aide of (\ref{trans1}) the above expression can be written in the representation of the delocalized atomic orbitals $\{ n \kk \}$
\begin{equation}  \label{ro3}
 \hat{\varrho}(\q) = \sum_{\kk \s}\sum_{n n'} eG^{n n'}_{\kk, \kk+\q} c^{\dag}_{n \kk \s} c^{\ph}_{n' \kk+ \q \s},
\end{equation}
with  $G^{n n'}_{\kk, \kk+\q} \approx \delta_{n n'}$. 
In a similar way, the charge density operator $\hat{\varrho}(\q)$ can be defined
in the  Bloch representation, using transformation (\ref{trans3})
\begin{eqnarray}  \label{ro4}
 \hat{\varrho}(\q) &=& \sum_{\kk \s}\sum_{L L'}e G^{L L'}_{\kk, \kk+\q} c^{\dag}_{L \kk \s} c^{\ph}_{L' \kk+ \q \s}  \nonumber \\
 &=&  \sum_{\kk \s}\sum_{L L'}e G^{L L'}_{\kk, \kk+\q}  \hat{\varrho}^{LL'}_{ \kk\s, \kk + \q\s}.
\end{eqnarray}
The operator $\hat{\varrho}^{LL'}_{ \kk\s, \kk + \q\s}$ is called the electron-hole propagator and it plays a pivotal role in the equations describing the charge density excitations.
The charge  vertex  $G^{L L'}_{\kk, \kk+\q}$ is given by 
\begin{equation}  \label{ro5}
 G^{L L'}_{\kk, \kk+\q} = \sum_{n n'} G^{n n'}_{\kk, \kk+\q} V_{\kk}(n , L) V^{*}_{ \kk+ \q}(n', L')
\end{equation}
and its explicit form is derived in Appendix \ref{appC}. These matrix elements are a trivial simplification of the matrix 
elements (\ref{Matrel}), obtained by setting $g_z=0$ and restricting the number of Bloch bands to two.

\section{Equation of motion for the electron-hole propagator}\label{jedbe}

Here an analysis is presented of the charge density fluctuations in a electron subsystem described by a single particle Hamiltonian (\ref{dijag}) to which a long-range electron-electron interaction is added.
Therefore in the context of the expression (\ref{ro4}), it is clear that the dynamics of the 
electron-hole propagator $\hat{\varrho}^{LL'}_{ \kk, \kk + \q} $ has to be determined in the presence of the $v$ and $c$ bands only, since at the point of interest (K point of the Brillouin zone)
the other bands are far enough away in energy (Fig.\ref{f1}(b)).
Hereafter, the spin index in the electron-hole propagator, whose
 dynamic is described by the Heisenberg equation
\begin{equation} \label{tri1}
i\hbar \frac{\partial }{\partial t} \hat{\varrho}^{vc}_{ \kk, \kk + \q}  =  \left[ \hat{\varrho}^{vc}_{ \kk, \kk + \q} , \hat{H}  \right] ,
\end{equation}
is omitted.

\begin{figure}[tt]
\includegraphics[width=8cm]{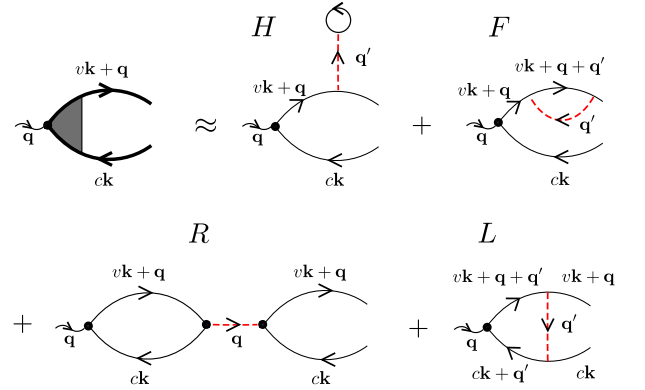} 
\caption{\small{Faynman diagrams of four first order contributions to the interband electron-hole propagator: Hartree (H), Fock (F), RPA (R) and ladder (L). Coulomb
interaction is represented by the red dashed line with $\q' = \kk'-\kk$.}}
\label{f2}
\end{figure}

The  Hamiltonian  in (\ref{tri1}) consists of the bare Hamiltonian (\ref{dijag}) 
and the Coulomb interaction term
\begin{eqnarray}\label{tri2}
\hspace{-10mm}  \hat{H}_{e-e} && = \frac{1}{2} \sum_{\q \neq 0} v_{\q} \hat{\varrho}^{\dag}(\q) \hat{\varrho}(\q) \nonumber \\
&&= \frac{1}{2V} \hspace{-2mm} \mathop{\sum_{\kk',\kk, \q,\s, \s'}}_{ L_1,L_2,L_3, L_4} \hspace{-3mm}  \mathbb{W}\bigl(\begin{smallmatrix}
L_1 & L_2 & L_3 & L_4 \\
\kk&\kk'&\kk'+\q&\kk - \q
\end{smallmatrix}\bigr)  \times \nonumber \\
&&c_{L_1 \kk \s}^{\dag}  c_{L_2 \kk' \s'}^{\dag}  c^{\ph}_{L_3 \kk'+\q \s'}  c^{\ph}_{L_4 \kk-\q \s}.
\end{eqnarray}
The two-particle Coulomb matrix elements in (\ref{tri2}) are given in terms of charge vertices  (\ref{ro5})
\begin{equation} \label{tri3}
\mathbb{W}\bigl(\begin{smallmatrix}
L_1 & L_2 & L_3 & L_4 \\
\kk&\kk'&\kk'+\q&\kk - \q
\end{smallmatrix}\bigr)    = v_{\q}e^2G^{L_1 L_4}_{\kk,\kk - \q} G^{L_2 L_3}_{\kk',\kk' + \q}
\end{equation}
with  $v_{\q} = 2\pi/|\q|$ being the Fourier transform of the bare Coulomb interaction in two 
dimensions. In the following sections, $v_{\q}$ will be replaced by screened Coulomb interaction.
The Hartree-Fock corrections to the single particle energies $\e_{\kk}^{L}$, $L \in \{v,c\}$ are introduced in the following way
\begin{equation}\label{tri4}
E^{L}_{ \kk} = \e_{\kk}^{L} + \sum_{\kk' L'}  \left[ 2\mathbb{W}\bigl(\begin{smallmatrix} L'& L& L& L' \\  \kk'&\kk&\kk&\kk'  \end{smallmatrix}\bigr)-
\mathbb{W}\bigl(\begin{smallmatrix} L'& L & L'& L \\  \kk'&\kk&\kk'&\kk  \end{smallmatrix}\bigr)\right] f^{L'}_{\kk'}.
\end{equation}
As already noted (Sec. \ref{parametri}), Hartree-Fock corrections have been  phenomenologically included in  the TBA dispersions. Therefore,  $E^{L}_{ \kk} = \e_{\kk}^{L}$.
The solution of the  equation (\ref{tri1}), evaluated using the Wick theorem \cite{mahan}, can  be written down to the first order in Coulomb interaction 
\begin{eqnarray}\label{tri5}
&&\hspace{-0mm} \left( \h\omega + E^v_{\kk} -E^c_{ \kk+\q}   \right) \hat{\varrho}^{vc}_{ \kk, \kk + \q}  = \nonumber \\
&&\hspace{-0mm}  \frac{2}{V} \sum_{\kk'} v_{\q}e^2G^{cv}_{\kk+\q,\kk}G^{vc}_{\kk',\kk'+\q} \left[ f^{v}_{\kk} - f^{c}_{\kk+\q}  \right] \hat{\varrho}^{vc}_{ \kk', \kk' + \q}   \nonumber \\
&&\hspace{-0mm} +\frac{1}{V}\sum_{\kk'} v_{\kk'-\p}e^2G^{cc}_{\kk+\q,\kk'+\q}G^{vv}_{\kk',\kk} \left[ f^{c}_{\kk+\q} - f^{v}_{\kk}  \right] \hat{\varrho}^{vc}_{ \kk', \kk' + \q}  . \nonumber \\
\end{eqnarray}
Within the  self-consistent equation (\ref{tri5}), the four main contributions to the interband electron-hole propagator can be defined. The Hartree and  Fock terms have been absorbed in the 
single particle energies, leaving RPA  and ladder contribution on the right side, respectively. 
These four contributions are depicted using Feynman diagrams in Fig.\ref{f2}.
It should be noted that (off-resonance) band changing scattering processes are omitted in Eq.\ref{tri5}. 
This, so called Tamm-Dancoff approximation \cite{T-D} is valid here, due to the $\q \approx 0 $ form of interband the intraband charge verticies, as shown 
in Appendix \ref{appC}. 

\section{Solutions of the (\ref{tri5}) in the  $\q \approx 0 $ limit } \label{aproksimacije}
The expression (\ref{tri5})  is a self-consistent integral equation which is usually solved  numerically.
Various approximations have to be performed in order to obtain a more appealing analytical solution.
The first step is to take the $\q \approx 0$ limit (since the lowest form of bound exciton energy are of interest), thus neglecting the kinetic energy of the electron-hole pair. 
In this limit the RPA contribution in the equation (\ref{tri5}) vanishes. This is because the interband charge vertices are proportional 
to $\q$ (Appendix \ref{appC}). Then even in the case of the bare Coulomb potential ($v_{\q} = 2\pi/|\q|$), the RPA term is proportional to $\q$, thus
leaving only the ladder part as the dominant term in (\ref{tri5}).
Next, interband transitions are restricted only to the narrow area around the K point between the fully occupied $v$ ($f^v_{\kk}=1$) and totally empty 
$c$ ($f^c_{\kk}=0$) band. At this point of direct transitions the Bloch dispersions are approximated by free electron dispersions with the 
effective masses. Redefining $\kk$ respectively to the $\mb{K}=(4\pi/3a,0)$ vector ($\kk \to \mb{K} -\kk$) gives
\begin{equation}\label{cet1}
 E^{c}_{\p}  -  E^{v}_{\p}  \approx 2\Delta  + {\hbar^2\mb{k}^2}/{2\mu},
\end{equation}
with the  reduced mass $\mu = m^*_c m^*_v/(m^*_c+ m^*_v)$ and the band gap $2\Delta$. 
Defining the exciton energies $\Omega$ relatively to the bottom of the $c$ band by substitution  $\Omega = \hbar\omega  - 2\Delta$ in (\ref{tri5}), yields
\begin{equation}\label{cet2}
{\hspace{-0mm}}\left( \Omega   - {\hbar^2\mb{k}^2}/{2\mu}   \right) \hat{\varrho}^{vc}_{ \kk, \kk }   = -\frac{1}{V}\sum_{\kk'} v_{\kk'-\p}e^2G^{cc}_{\p,\kk'}G^{vv}_{\kk',\p}  \hat{\varrho}^{vc}_{ \kk', \kk'}  .
\end{equation}
Subsequently, the equation (\ref{cet2}) is solved in the  cases of bare and dynamically screened Coulomb potential.

\subsection{Wannier model} \label{wannier}
\begin{figure*}[t]
\includegraphics[width=14cm]{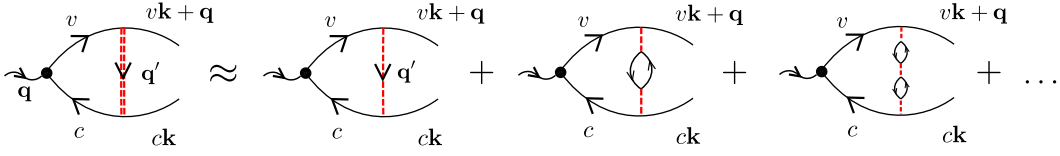} 
\caption{\small{Schematic representation of the  RPA contributions to the  ladder diagram of the electron-hole propagator.
The bare and the screened Coulomb interactions with $\q' = \kk'-\kk$ are denoted by red dashed and double red dashed lines, respectively.}}
\label{f3}
\end{figure*}

In this section, Eq. (\ref{cet2}) is solved for the case of the bare Coulomb potential $v_{\q'} = {2\pi}/{|\q'|} $  where $\q' = \kk' - \kk$,
in a similar way as it was done by  Wannier \cite{wannier}. 
Due to the singular
 behavior of the Coulomb interaction, the main contribution in the sum on the right hand side of (\ref{cet2}) comes from the $\q' \approx 0$ states. 
Also, considering that the intraband charge verticies are equal to one, leads to
\begin{equation} \label{sex1}
\left( \Omega   - {\hbar^2\mb{k}^2}/{2\mu}   \right) \hat{\varrho}^{vc}_{ \kk, \kk }  \approx -\frac{1}{V}\sum_{\q' \approx 0} \frac{2\pi e^2}{|\q'|} \hat{\varrho}^{vc}_{ \kk+\q', \kk+\q' }.
\end{equation}
This  self-consistent equation can be Fourier transformed into the direct space representation, by introducing the Fourier transform of the electron-hole 
propagator amplitude
\begin{equation} \label{sex2}
\psi(\mb{r}) = \frac{1}{V}\sum_{\kk} \hat{\varrho}^{vc}_{ \kk, \kk } \m{e}^{i\kk \cd \rr },
\end{equation}
which is, by assumption, a smooth and well behaved function $(\psi( \infty)=\nabla \psi( \infty) =0 )$ of the electron-hole distance $|\mb{r}|$.  The  Eq. (\ref{sex1}) then  becomes
two-body Schr\"{o}dinger-like equation 
\begin{equation} \label{sex3}
\left[  -\frac{\hbar^2}{2\mu}\nabla^2 - \frac{e^2}{|\rr|}   \right] \psi(\rr) = \Omega \psi(\rr).
\end{equation}
After the separation of variables  $\psi(\rr) = R(r)\Phi(\varphi)$,
two equations are obtained, which describe the radial and angular part of the wave function
\begin{eqnarray} \label{sex4}
&& \frac{\partial^2 R(r)}{\partial r ^2}  + \frac{1}{r} \frac{\partial R(r)}{\partial r }  + \frac{2\mu}{\hbar^2} \left( \Omega + \frac{e^2}{r}    \right) R(r) - \frac{\ell^2}{r^2}R(r) = 0 \nonumber \\
&& \ell^2 = -\frac{1}{\Phi(\varphi)} \frac{\partial^2 \Phi(\varphi)}{\partial \varphi ^2}.
\end{eqnarray}
The eigenvalues are well-known  2D hydrogen-like form \cite{vodik}
\begin{equation} \label{sex5}
\Omega_n =  - \frac{\mu}{m_e}\frac{1 \,  \rm{Ry}}{(n-1/2)^2},
\end{equation}
with  each state $n$
being $(n-1)$ times degenerated (not including the valley degeneracy which brings an extra factor of 2), since $\ell = 0,1,\cdots n-1$.
By inserting the reduced mass $\mu = 0.35m_e$ (\ref{effe2}) in the above equation, the ground state exciton energy in 
hBN-ML $\Omega_1 =-1.4\,$Ry ($-19$ eV) is obtained. It is evident that $\Omega_1$
is too low, suggesting that Wannier model should  be improved by the introduction of the 
screened Coulomb interaction.

\subsection{Screening of Coulomb interaction}

It is shown (Eq.\ref{tri5}) that to the first order in  $\mathcal{O}(v_{\q'})$, the ladder term is the one describing the dynamics of the electron-hole pair in the $\q \approx 0$ limit.
The inclusion of the higher order contributions ($\mathcal{O}(v^n_{\q'})$) is usually performed by summing infinite number of RPA contributions  to the  ladder diagram of the electron-hole propagator, shown by Feynman diagrams in Fig.\ref{f3}.
This procedure  is equivalent to 
changing the bare Coulomb interaction to  a screened one in the expression (\ref{sex1})
\begin{equation} \label{sed1}
v_{\q'} \to {w}_{\q'}(\om) = v_{\q'}/ \epsilon(\q',\om).
\end{equation}
 The RPA longitudinal dielectric function is 
\begin{equation}\label{sed2}
 \epsilon(\q',\om) = 1-v_{\q'}\chi^0(\q',\om), 
\end{equation}
where the density-density response function, known as the generalized Lindhard function 
\cite{mahan} can be written as
\begin{equation}\label{sed3}
\chi^0(\q',\om) = \frac{e^2}{V}\sum_{LL'\kk \s} \left| G^{LL'}_{\kk,\kk+\q'} \right|^2
\frac{f^L_{ \kk} - f^{L'}_{\kk + \q'}}{\hbar\om + E_{ \kk}^L - E_{ \kk+ \q'}^{L'} + i\eta} .
\end{equation}
Here, Eq. (\ref{sed2}) is evaluated within the two-band TBA model presented in section \ref{TBA} for the  $\q' \approx 0$. The Hartree-Fock energies are 
given by the TBA energies $E_{ \kk}^L = \ee^L_{\kk}$  (considering that many 
particle correlations are already implemented in the simple single particle TBA model) and  the charge vertices $G^{LL'}_{\kk,\kk+\q'}$ are derived in Appendix \ref{appC}.
Finally, to the first order in $\mathcal{O}(\q')$, the dielectric function is given by
\begin{equation}\label{sed4}
\epsilon(\q',\om) =  1+2\pi \al(\om) |\q'| + \mathcal{O}(\kappa^2).
\end{equation}
The function $\al(\om)$ is called the electronic polarizability of the insulator. It is usually calculated  
{\it ab initio}  \cite{vito1}, however, here it is evaluated  within the two-band model presented
earlier using the relaxation constant approximation. In this case  the adiabatic term $\eta$ in (\ref{sed3}) is replaced by an interband relaxation constant $\Gamma$.
The results for $\al(\om)$ for various relaxation constants $\Gamma$ are presented in the Fig.\ref{f4}(a).

In the case of  $\eta \to 0$, the polarizability function $\al(\om)$ can be analytically estimated by considering the direct transitions around the K point. 
The expression (\ref{sed3}) is  integrated to the cut-off wave vector  $k_0$  which determines the extend around the K point to which the bands are described by a parabolic dispersion
with the effective masses (\ref{effe2}).
By inspecting the band structure in Fig.\ref{f1}(b) the value of $k_0 \approx {K}/3$ can be chosen. The result is 
\begin{eqnarray}\label{sed5}
&& \Re \al(\om)  =  \frac{e^2}{2\pi} \frac{1}{2\Delta}{ \rm{ln} }\frac{(\hbar\om)^2 - \left(2\D + \e_0  \right)^2}{(\hbar \om)^2 - (2\D)^2} \nonumber \\
&& \Im \al(\om)  = 0.
\end{eqnarray}
Here, the cut-off energy is  $\ee_0 = \h^2k_0^2/(2m^*_c) \approx 3$ eV.  The $\alpha(\om)$ given by  (\ref{sed5}) is in excellent agreement with the exact two-band TBA 
model result for vanishing $\Gamma=1\,$meV, as shown by the brown circles in the Fig.\ref{f4}(a).
Inserting (\ref{sed4}) into (\ref{sed1}) the dynamically screened Coulomb potential 
\begin{equation} \label{os1}
 w_{\q'}(\om) = \frac{2\pi}{|\q'|(1+2\pi \al(\om) |\q'|)}
\end{equation}
is obtained. 

 After Fourier transformation to the direct space it 
becomes a $\omega$-dependent Keldysh potential \cite{keldysh,cudazzo, vignale}
\begin{equation} \label{os2}
 W(r,\om) = \frac{e^2 \pi}{2\ld(\om)} \left[Y_0(r/\ld(\om)) - N_0 (r/\ld(\om)) \right],
\end{equation}
where $Y_0(x)$  and $N_0(x)$ represents Struve and  Neumann function respectively. The dynamical screening 
length is $\ld(\om)=2\pi\al(\om)$, where $\al(\om)$ is given by (\ref{sed5}). Next, 
the bare Coulomb potential ${e^2}/{r}$ is replaced in equation (\ref{sex4}) by $W(r,\om)$ and solved self-consistently.

\section{Results and discussion}

\subsection{Exciton energies  and  spatial extent}\label{rez} 

 \begin{figure*}[tt]
\includegraphics[width=8.3cm]{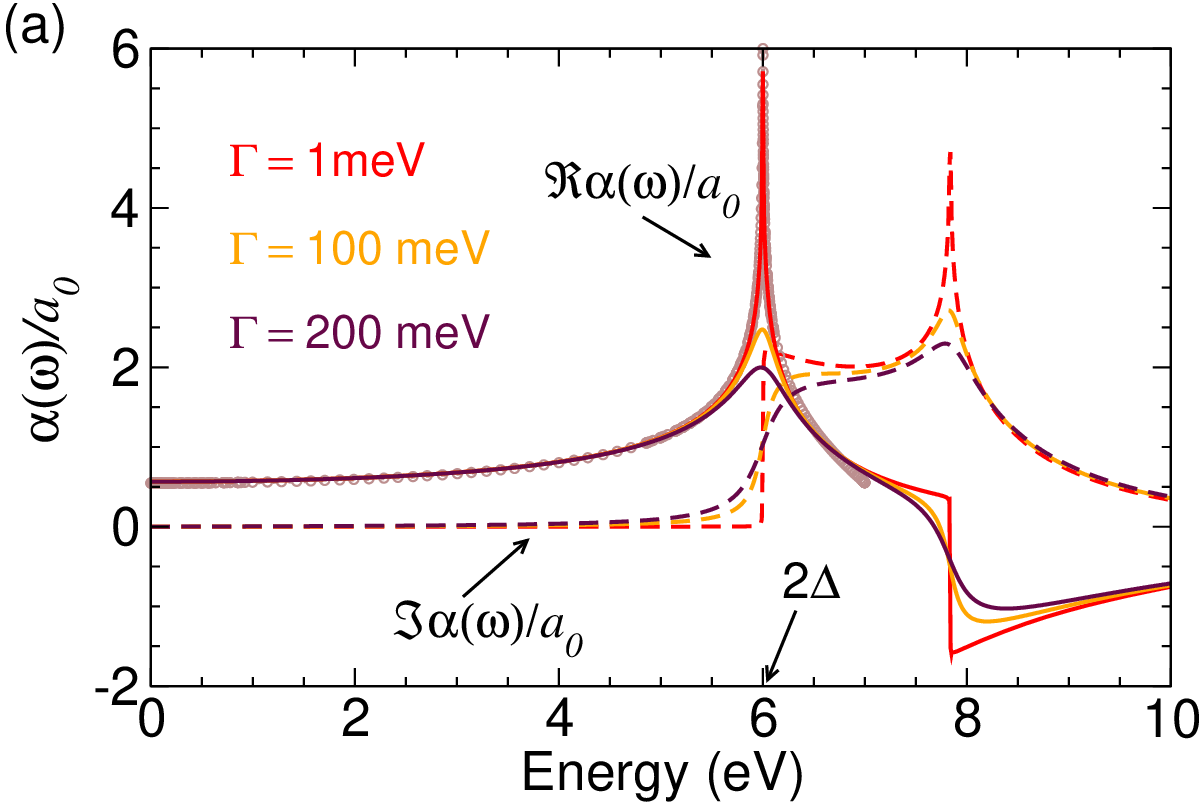} 
\hspace{0.8cm}
\includegraphics[width=8.1cm]{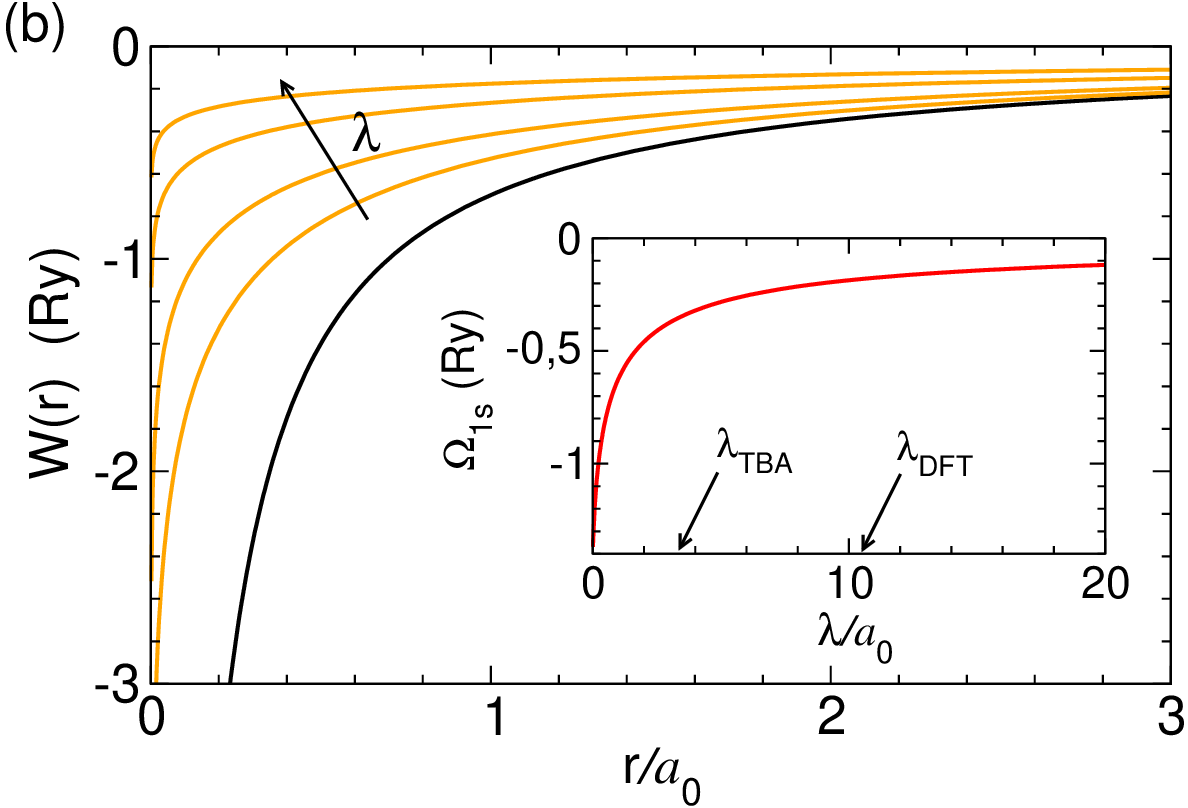}
\caption{\small{(a) Real (solid line) and imaginary (dashed line) parts of the dynamical polarizability $\al(\om)$ of hBN-ML, calculated using the two-band TBA 
model for different values of the interband relaxation rate $\Gamma$. The brown circles represent the approximate expression (\ref{sed5}).
(b)  The bare Coulomb potential (black) and the screened Coulomb potential (\ref{os2}) (orange) plotted for different parameters $\ld = 1, 2,5,10\ a_0$. 
The insert shows dependence of the exciton ground state energy $\Omega_{1s}$ on $\ld$ for $\mu = 0.35m_e$.}  }
\label{f4}
\end{figure*}

Here the solutions of the Schr\"{o}dinger equation (\ref{sex4}) with the potential (\ref{os2}) are presented. 
The  exciton energies are given in Table \ref{tabla1} and are presented in terms of two quantum numbers $(n_r,\ell)$. The radial quantum number $n_r$ 
gives the number of nodes in the radial part of the wave function $R(r)$ and $\ell$ is the orbital quantum number.  
The states denoted by the  $\ell =0,1,2,...$ are labeled as $s,p,d,..$. This is an adequate choice since the absolute square of the angular part of the 
wave function behaves like $|\Phi(\varphi)|^2 \sim \rm{cos}^2(\ell \varphi)$, thus resembling to the 2D projections of the 3D atomic hydrogen orbitals. Moreover, 
the principal quantum number $n = 1 + n_r + \ell $ can be introduced. Then the energy states labeled as $\Omega(n_r,\ell)$ can be equally labeled $\Omega_{n\ell}$. For example, a state 
 $\Omega(n_r=1,\ell=1)$ is equivalent to the $\Omega_{3p}$ state, etc. All energy states  having the same $n$ are given in the same color in 
Table \ref{tabla1}.
 \begin{table}[!ht] 
\begin{tabular}{|C{1.3cm}||C{1.6cm}|C{1.6cm}|C{1.6cm}|C{1.6cm}|}
\hline
$|\Omega(n_r,\ell)|$ & $\ell = 0$ & $\ell = 1$ & $\ell=2$ & $\ell=3$  \\ \hline\hline
$n_r = 0$ & \textcolor{maroon}{4.64 (4.73 )} & \textcolor{red}{ 1.30 (1.57)} & \textcolor{orange}{0.60 (0.70)} & \textcolor{gold}{0.34 (0.38)}  \\ \hline
$n_r =1$ & \textcolor{red}{0.95 (1.20)} & \textcolor{orange}{0.52 (0.64)} & \textcolor{gold}{ 0.31 (0.37) }    \\ \cline{1-4}
$n_r = 2$ & \textcolor{orange}{0.42 (0.54)} & \textcolor{gold}{ 0.28 (0.34)}    \\ \cline{1-3}
$n_r = 3$ & \textcolor{gold}{0.24 (0.30)}    \\ \cline{1-2}
\end{tabular}
\caption{(color online) The first ten exciton energy levels in eV. The states within the same shell are given in the same color: $n=1,2,3,4$ in maroon, red, orange and 
olive green, respectively.}
\label{tabla1}
\end{table}
The first entry in Table \ref{tabla1} is the  exciton energy obtained by solving the self-consistently equation (\ref{sex4}) 
with the frequency depended potential $W(r,\om)$. The second entry (in the brackets) is the solution of the (\ref{sex4})  in the case of the static
potential $W(r,0)$ where expression (\ref{sed5}) was used to calculate $\ld(0)= \ld_{TBA} = 3.5 \, a_0$.    
By examining the $\Omega_{1s}$ state from the Table \ref{tabla1}, it can be seen that even small (two-band 
TBA model) values of $\ld(\om)$  produce a  strong reduction of the exciton ground state energy in comparison with the  energy of $-19\,$eV obtained using the bare Coulomb 
potential, i.e. from Eq. (\ref{sex5}).

\begin{figure}[t]
\includegraphics[width=8.2cm]{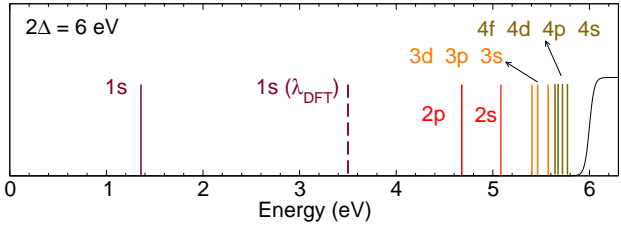} 
\caption{\small{The excitonic spectrum of hBN-ML containing first ten levels (solid lines) from the Table \ref{tabla1}. The dashed maroon line represents the  
exciton  ground state energy $\Omega_{1s}(\ld_{DFT}) = -2.5\,$eV calculated using the $\ld_{DFT}$ static screening length.
 The threshold energy for the single particle excitations is represented by a black 
line. }}
\label{f6}
\end{figure}

The states with higher $\ell$ have lower energy for the same $n$.  This can be  seen from the Fig.\ref{f6} in the case of 
$n =3$ series. The energy ordering of the states $\Omega_{3d} < \Omega_{3p} < \Omega_{3s}$ is the same, regardless whether they are calculated with dynamical or statical Keldysh potential. This energy ordering is 
experimentally observed in the two-photon absorption experiments on tungsten disulphide \cite{exi-naj}. 
The relative  difference between $\Omega_{n\ell}$ obtained by the $W(r,\om)$ and those obtained by the $W(r,0)$  increases as the band gap edge is approached. 
For  $\hbar\omega \ll 2\Delta$  the Keldysh potential can be approximated by its statical limit since $\ld(\hbar\omega \ll 2\Delta) \approx \ld(0)$ Fig.\ref{f4}(a).
As the conduction band is approached the logarithmic divergence in the dynamical screening length $\ld(\om)$ becomes more apparent and $\Omega_{n\ell}$ increase compared with those 
calculated with $\ld(0)$.
The mean exciton radius,  defined as the average electron-hole separation in the state 
$\psi_{n\ell}$,  is calculated as $\overline{r}_{n\ell} = \la \psi_{n\ell} |r| \psi_{n\ell}  \ra$.
In the ground state $\overline{r}_{1s} \approx 4 \, a_0$, which is comparable with 
the unit cell dimension. 
However, it should be noted that the two-band TBA polarizabilities are low in comparison with  $\ld_{{\it DFT}} = 10.5 \,a_0$ giving $\overline{r}_{1s} \approx 10 \, a_0$,
which is in accordance with the Wannier scheme. The mean exciton radius decreases with $\Omega_{n \ell}$. 
For example, $\overline{r}_{2s} \approx 20\, a_0$ and $\overline{r}_{2p} \approx 13\, a_0$, while for 
the highest calculated energy level $\overline{r}_{4s} \approx 83\, a_0$.

In the Fig.\ref{f4}(b)   the screened  potential $W(r,0)$ is plotted as a 
function of the  screening length $\ld$. As $\ld$  increases, the logarithmic nature of the potential becomes 
more apparent for small electron-hole separations. 
This can be seen  by taking the two opposite limits of the expression (\ref{os2})
\begin{eqnarray}  
&& W(r\to \infty,0) \to -e^2/r \label{os31}  \\
&& W(r \to 0,0) \to (e^2/\ld)\m{ln}(r{\rm{e}}^{\gamma}/2\ld), \label{os32} 
\end{eqnarray}
where $\gamma$ is Euler-Mascherion constant.

The shape of the above potential suggests the reason why the states within the same shell $n$ differ in energy.
The radial wave function $R(r)$ of the states with lower $\ell$ is located closer to the origin where it is governed by a weaker logarithmic potential (\ref{os32}).
Therefore, the binding energy is smaller and the mean radius is larger compared to the states with larger values of $\ell$ which  are under the influence of 
stronger bare Coulomb potential (\ref{os31}) and thus have larger binding energy and smaller spatial extension. 
It is useful to determine the dependence of the  exciton ground state energy $\Omega_{1s}$ as a function
of the screening length $\ld$. This dependence is shown in the insert of Fig.\ref{f4}(b).
In the limit $\ld \to 0$,  $\Omega_{1s}(\ld)$ is given by the expression (\ref{sex5}), while in the opposite limit $(\ld \to \infty)$ 
a saturation of the $\Omega_{1s}(\ld)$ can be  seen. The analytical approximation of $\Omega_{1s}(\ld \to \infty)$ will be 
considered in the last section.

\subsection{The single particle gap problem}\label{gap}

In optical absorption experiments on quasi hBN-ML \cite{gap-broj,piret} the  exciton ground state signal appears at energy $\hbar \om_{1s}^{exp} \approx6$ eV. 
The theoretical result for the exciton ground state energy, calculated using the 
screened model, with {\it ab initio}  screening length $\ld_{DFT} = 10.5 \,a_0$, yields $\hbar\om_{1s} = 2\D + \Omega_{1s}(\ld_{DFT}) = 3.5$ eV. 
This result shows that the calculated exciton energies would  agree well with those obtained in the  
absorption experiments only if the quasiparticle band gap were larger then the value given by the LDA-DFT calculations. 
This is not surprising considering 
that the approximation used within DFT calculations do not  take the many-particle correlation effects 
properly and  the single particle band gap is usually systematically underestimated.
Using the results of Sec.\ref{rez} the lower limit of the hBN-ML band gap can be estimated to be about $9$ eV. 
This can be done by searching for the value of the band gap $2\Delta'$ for which the theoretically obtained exciton energy $\hbar\om_{1s}$ is equal to the experimental 
value $\hbar\om_{1s}^{exp}$.
Here it should be taken into consideration that the static screening lengths $\ld_{DFT}$ also depend on the band gap, decreasing as the band gap increases. 
However, the exciton binding energy $\Omega_{1s}$ (as shown in Fig.4(b) insert) depends relatively weakly on $\ld$  for  $\ld \approx \ld_{DFT}$, so the same value $\Omega_{1s}(\ld_{DFT})$ can be used.
Therefore, from  
\begin{equation} \label{exc-exp}
\hbar\om_{1s}^{exp} =  2\Delta' + \Omega_{1s}(\ld_{DFT}),
\end{equation}
a lower limit of the quasiparticle band gap is estimated as 
$2\Delta' \geq 8.5$ eV. This simple estimation agrees well with the statical COH-SEX correction of band gap, $2\Delta_{GW}=9$ eV, as presented in 
Sec. \ref{abin}.

Observing Fig.\ref{f6}, some conclusions can be made regarding the interaction  of excitons with phonons, impurities, and electrons, changing their appearance in the absorption 
spectrum. Due to these interactions, the exciton signals will be broader and slightly shifted in energy. This causes overlapping between the individual exciton signals 
that are close in energy, to the point that they can even be joined with the single particle threshold.
This would imply the indistinguishability of partial contributions originating from the single particle  excitations and the excitons in the optical absorption spectra.

\subsection{The limit of high polarizability - EBK procedure}
\begin{figure}[tt]
\includegraphics[width=8.3cm]{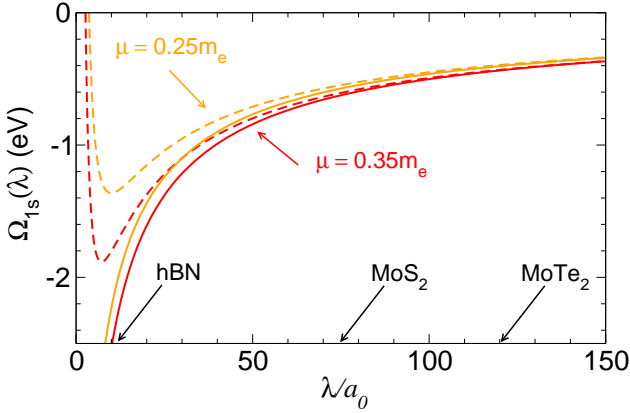} 
\caption{\small{Exciton ground state energy $\Omega_{1s}$ as a function of the screening length $\ld$ obtained by solving 
the Schr\"{o}dinger equation for the statical screened potential  (\ref{os2}) (solid lines) and using the  analytical expression  Eq. (\ref{wkb3}) 
(dashed lines) with $z=3.5$ for two values of the reduced masses $\mu=0.35m_e$ (red) and $\mu=0.25m_e$ (orange).
The {\it ab initio} values of the static screening lengths of hBN, $\rm{MoS_2}$ and $\rm{MoTe_2}$ monolayers are indicated.}}
\label{f5}
\end{figure}

The analytical form of the ground state exciton energy $\Omega_{1s}$ is presented in the limit of high screening length
 $(\ld  \to \infty)$. As the  {\it ab initio} results predict, hBN-ML can not be considered a material where the above condition 
applies, since $\ld_{DFT} = 10.5 \, a_0$. However, other two-dimensional materials, such as transition-metal dichalcogenides, have much larger $\ld$. 
DFT calculations for molybdenum disulfide ($\rm{MoS_2}$-ML) and  molybdenum ditelluride ($\rm{MoTe_2}$-ML) were also performed, giving
$\ld_{DFT} = 76 \ a_0$ and $120 \ a_0$ and $\mu = 0.25m_e$ and $0.3m_e$, respectively. 
This is a direct  consequence of their spatial structure where a transition metal plane lies between the planes of two chalcogenide atoms.
This geometrical coordination reduces the electronic hopping elements between the atomic orbitals of the neighboring atoms \cite{vito2} and hence 
causes flattering of electron bands. Smaller dispersivity of the electron bands
combined with the smaller band gap, such as  $2\Delta_{DFT} \approx 1.8$ eV in the case of $\rm{MoS_2}$-ML and $2\Delta_{DFT} \approx 1.2$ eV in the case 
of $\rm{MoTe_2}$-ML, give almost an order of magnitude larger static screening length in comparison with $\ld_{DFT}=10.5 \, a_0$ for hBN-ML.
This will certainly be responsible for the logarithmic behavior of the screened potential over sizable electron-hole spatial extension, as can be 
seen in Fig.\ref{f4}(b), making the logarithmic potential (\ref{os32}) adequate for determining the exciton ground state energy.

In this case  the semiclasical EBK approach can be applied to calculate the 
low-lying ($\ell = 0$) exiton energy levels. The general EBK approximation \cite{wkb} asserts that in the spherical symmetric problem the phase integral of 
the radial impulse is quantized as
\begin{equation} \label{wkb1}
 2 \int_0^{r_0} \sqrt{2\mu \left[ \Omega - W(r\to 0,0) \right]} dr = \pi \hbar(n_r + z/4).
\end{equation}
Here  $r_0$ is the classical turning point,  $n_r=0$ is the radial quantum number for the ground state energy case,
and $z=2$ is the Maslov index  \cite{wkb},  which gives the number of classical turning points. This leads to the  implicit expression for the exciton ground state
energy
\begin{equation}\label{wkb2}
\frac{\sqrt{\pi}}{2}  {\rm{Erf}} \left( \sqrt{{\Omega \ld}/{e^2}}  \right){\rm{e}}^{{\Omega \ld}/{e^2}} - \sqrt{{\Omega \ld}/{e^2}}  = \frac{\pi \hbar {\rm{e}}^{\gamma} {z}/{4}}{ \sqrt{32\mu e^2\ld }}.
\end{equation}
In the case of large screening length ($\ld/e^2 \to \infty$),  when the exponential function dominates and 
the error function  is ${\rm{Erf}}(x \to \infty)=1$, Eq. (\ref{wkb2}) can be simplified, which finally leads to 
the analytical expression for the exciton ground state energy
\begin{equation}\label{wkb3}
\Omega_{1s}(\ld) \approx \frac{e^2}{\ld}  {\rm{ln}} \left( \frac{\hbar{\rm{e}}^{\gamma} z \sqrt{\pi}}{8\sqrt{2\ld \mu e^2}}  \right).
\end{equation}
The similar expression has been found as the limiting solution of the Schr\"{o}dinger equation for the logarithmic potential  \cite{ln-pot-sch}, unlike the results of \cite{prl1} which 
predict $\Omega_{1s}(\ld) \approx -(3/4\pi)e^2/\ld$ in this limit and thus neglecting the logarithmic term. 
The expression (\ref{wkb3}) gives a fairly accurate description of $\Omega_{1s}(\ld)$ in $\ld \to \infty$ limit, but it can be improved by setting $z=3.5$.
Fig.\ref{f5} shows the comparison between the ground state  energy $\Omega_{1s}(\ld)$ obtained using the static screened potential   (\ref{os2}) (solid lines)
and using the analytical form (\ref{wkb3}) with $z=3.5$ (dashed lines), for two  different reduced mases $\mu$.
It is evident that the agreement between the curves becomes noticeable for large values of $\ld$.

Using the same arguments outlined in Sec. \ref{gap}, together with Eq. (\ref{wkb3}), the single particle band gap of  
$\rm{MoS_2}$ and $\rm{MoTe_2}$ monolayers  can be  estimated,
in which the experimentally determined exciton energies are $\hbar \om_{1s}^{exp}(\rm{MoS_2}) \approx 1.9 $ eV 
and $\hbar \om_{1s}^{exp}(\rm{MoTe_2}) \approx 1.2$ eV \cite{mos-exc,mos-exc2, mote-exc}.
Inserting the calculated polarizabilities and reduced masses into (\ref{wkb3}) and using (\ref{exc-exp}) gives $2\Delta'_{\rm{MoS_2}} \approx 2.5$ eV and 
$2\Delta'_{\rm{MoTe_2}} \approx 1.6$ eV. The DFT polarizabilities, the experimental exciton energies and estimated quasiparticle band-gaps for three studied 2D crystals are sumarised 
in Table \ref{tabla2}.
\begin{table}[!ht] 
\begin{tabular}{|C{1.5cm}||C{1.6cm}|C{1.6cm}|C{1.6cm}|}
\hline
2D cryst  & hBN & MoS$_2$ & MoTe$_2$  \\ \hline\hline
$\lambda_{DFT}/a_0$  & $10.5$ & $76$ & $120$  \\ \hline
$\mu/m_e$  & $0.35$ & $0.25$ & $0.3$  \\ \hline
$\hbar \om_{1s}^{exp}/eV$& $6.0$ & $1.9$ & $1.2$  \\ \hline
$2\Delta'/eV$  & $8.5$  & $2.5$  & $1.6$  \\ \hline
\end{tabular} 
\caption{The quasi-particle band-gap $2\Delta'$ estimated from DFT  polarizabilities $\lambda_{DFT}$, reduced masses $\mu$,  the experimental exciton energies $\hbar \om_{1s}^{exp}$ and from 
Eqs. (\ref{exc-exp}) and (\ref{wkb3}).}
\label{tabla2}
\end{table}
The calculated values of the singleparticle band gaps are in accordance with the results of the BSE-GW approach\cite{gapovi1,gapovi2,gapovi3}.

\section{Conclusion}
By examining the equation of motion for the electron-hole propagator, in the leading order of Coulomb interaction, it was possible to reduce it to the two-body Schr\"{o}dinger 
equation. The inclusion of the higher order contributions to the electron-hole propagator leads to the Schr\"{o}dinger equation with the dynamicaly screened 
Coulomb potential. The parameters like the dynamical screening length
and the electron and hole effective masses are obtained from the TBA approximation  and supported by {\em ab initio} calculations. 
 Using this parameters the exciton binding energies and wave functions are calculated. The
exciton binding energies obtained using the bare Coulomb potential are compared to the ones obtained using the staticaly screened Coulomb potential, demonstrating how the bare Coulomb interaction leads to the unrealistically large exciton binding energy.
Comparing the calculated and experimental exciton ground state energies, a realistic value of the single particle band gap is estimated, which in the case 
of hBN-ML is $8.5$ eV. Using the EBK procedure an analytical expression for the exciton ground state energy is obtained in the high polarizability limit. This is shown to be valid for the 
family of transition metal dichalcogenides where the single particle band gap is calculated for  $\rm{MoS_2}$ and $\rm{MoTe_2}$ monolayers.

\begin{acknowledgements}
This work was supported by the QuantiXLie Centre of Excellence, a project
cofinanced by the Croatian Government and European Union through the
European Regional Development Fund - the Competitiveness and Cohesion
Operational Programme (Grant KK.01.1.1.01.0004). The authors would like to thank Ivan Kup\v{c}i\'c and  Kre\v{s}imir Cindri\'c for many  stimulating  discussions.
\end{acknowledgements}

\begin{appendix}

\section{matrices U and V} \label{appA}

Bloch operators $c^{\dag}_{L \kk \s}$ are a solution to the Heisenberg equation
\begin{equation} \label{a1}
  [ \hat{H}_0, c^{\dag}_{L \kk \s} ] = \ee^L_{\kk}c^{\dag}_{L \kk \s}.
\end{equation}
Inserting (\ref{trans2}) in (\ref{a1}) one obtains the eigenvalue problem
\begin{equation} \label{a2}
 \sum_{n'} U_{\kk}(L,n') \left(  H^{nn'}_{\kk} - \e^L_{\kk}\delta_{nn'} \right) = 0,
\end{equation}
where the orthogonalization condition for  for Bloch functions implies the unitarity of matrix $\mb{U}$ 
\begin{equation} \label{a3}
 \big\{ c^{\ph}_{L \kk \s } , c^{\dag}_{L' \kk' \s} \big \}   = \delta_{LL'}\delta_{\kk \kk' }  \rightarrow \sum_{n} |U_{\kk}(L,n)|^2=1.
\end{equation}
Introducing the amplitude and the phase of the parameter $t_{\kk}$ (\ref{tk}) and the auxiliary phase $\vartheta_{\kk}$ 
\begin{equation} \label{a4}
  t_{\kk}  =|t_{\kk}|e^{i\varphi_{\kk}}, \hspace{3mm} \tan \varphi_{\kk} = \frac{\Im t_{\kk}}{\Re t_{\kk}},\hspace{3mm} \tan \vartheta_{\kk} = \frac{|t_{\kk}|}{\D},
\end{equation}
the matrix $\mb{U}$ can now be written
\begin{equation} \label{a5}
U_{\kk}(L,n) =  \begin{pmatrix}
  e^{-i\varphi_{\kk}}\cos ({\vartheta_{\kk}}/{2}) &  -\sin ({\vartheta_{\kk}}/{2}) 
  \vspace{2mm} \\
  e^{-i\varphi_{\kk}}\sin ({\vartheta_{\kk}}/{2}) &  \cos ({\vartheta_{\kk}}/{2})
 \end{pmatrix}.
\end{equation}
The inverse matrix $\mb{V}$ is then  
\begin{equation} \label{a6}
V_{\kk}(n,L) =  \begin{pmatrix}
  e^{i\varphi_{\kk}}\cos ({\vartheta_{\kk}}/{2}) & e^{i\varphi_{\kk}}\sin ({\vartheta_{\kk}}/{2})
  \vspace{2mm} \\
  - \sin ({\vartheta_{\kk}}/{2})  &  \cos ({\vartheta_{\kk}}/{2})
 \end{pmatrix}.
\end{equation}

\section{useful derivatives}\label{appB}

For explicit determination of the charge verticies (\ref{ro5}) the following derivatives have  to be determined
${ \partial |t_{\kk}|  }/{\partial k_{\alpha} }$,  ${ \partial \varphi_{\kk}  }/{\partial k_{\alpha}}$ and ${ \partial \vartheta_{\kk}  }/{\partial k_{\alpha}}$ where
$\alpha \in \{x,y \}$ is an Cartesian coordinate. 
From the definition (\ref{a4}) one has
\begin{equation} \label{b1}
   \frac{ \partial \vartheta_{\kk}  }{\partial k_{\alpha}}  = \frac{(1/\Delta){ \partial |t_{\kk}|  }/{\partial k_{\alpha}} }{1+ \tan^2 \vartheta_{\kk}},
\end{equation}
and the explicit derivatives of $|t_{\kk}|$ are
\begin{eqnarray} \label{b2}
   &&\frac{ \partial |t_{\kk}|  }{\partial k_{x} } = -\frac{t_0^2a\sqrt{3}}{|t_{\kk}|} \left(\sin ak_x + \sin \frac{ak_x}{2} \cos \frac{ak_y\sqrt{3}}{2} \right), \nonumber \\
 &&\frac{ \partial |t_{\kk}|  }{\partial k_{y} } = -\frac{t_0^2a\sqrt{3}}{|t_{\kk}|}\cos \frac{ak_x}{2} \sin \frac{ak_y\sqrt{3}}{2}.
\end{eqnarray}
Explicit derivatives of  $\varphi_{\kk} $ are
\begin{eqnarray}\label{b3}
 &&\frac{ \partial \varphi_{\kk}  }{\partial k_{y}} = -\frac{t_0^2a\sqrt{3}}{|t_{\kk}|} \left(-\cos ak_x + \cos \frac{ak_x}{2} \cos \frac{ak_y\sqrt{3}}{2} \right), \nonumber \\
 && \frac{ \partial \varphi_{\kk}  }{\partial k_{x}} = -\frac{t_0^2a\sqrt{3}}{|t_{\kk}|}\sin \frac{ak_x}{2} \sin \frac{ak_y\sqrt{3}}{2}.
\end{eqnarray}
In the Dirac regime, i.e. for the states ${\kk} \approx \mb{K}$, the derivatives of the amplitude and the phase of the hopping parameter 
simplify substantially. Introducing $\widetilde{\kk} =  \mb{K}- \kk$, gives 
\begin{eqnarray} \label{b4}
&& \frac{ \partial |t_{\tilde{\kk}}|  }{\partial k_{\al}} =  \frac{t_0a\sqrt{3}}{2}\frac{\tilde{k}_x\delta_{\al,x} + \tilde{k}_y\delta_{\al,y}}{\sqrt{\tilde{k}_x^2+\tilde{k}_y^2}} \nonumber \\
&& \frac{ \partial \varphi_{\tilde{\kk}}  }{\partial k_{\al}} = \frac{\tilde{k}_x\delta_{\al,y} - \tilde{k}_y\delta_{\al,x}}{\tilde{k}_x^2+ \tilde{k}_y^2}.
\end{eqnarray}

\vspace{0mm}
\section{charge vertices}\label{appC}
Explicit values of the interband and intraband charge verticies in the long wave limit are obtained by inserting the matrix elements (\ref{a6}) in (\ref{ro5}) and expand them in the leading order in $\q$.
The result is
\begin{eqnarray}  \label{c1}
 && G^{vc}_{ \kk, \kk+\q } = \left( G^{cv}_{ \kk, \kk+\q } \right)^*, \nonumber \\
 && G^{vc}_{ \kk, \kk+\q } \approx \frac{1}{2} \sum_{\al}q_{\al}\frac{\partial \vartheta_{\kk}}{\partial k_{\al}} - \frac{i}{2}\sin {\vartheta_{\kk}} \sum_{\al}q_{\al}\frac{\partial \varphi_{\kk}}{\partial k_{\al}}, \nonumber \\
 && G^{cc}_{ \kk, \kk+\q } =  G^{vv}_{ \kk, \kk+\q } \approx 1.
\end{eqnarray}
Around the $\mb{K}$ point this result simplifies. Inserting (\ref{b1}) and (\ref{b4}) in (\ref{c1}), leaves 
\begin{equation}  \label{c2}
 |G^{cc}_{ \tilde{\kk}, \tilde{\kk}+\q }| \approx 1, \hspace{4mm} |G^{vc}_{ \tilde{\kk}, \tilde{\kk}+\q }| \approx \frac{t_0a\sqrt{3}}{4\Delta}|\q|.
\end{equation}

\end{appendix}


\begin{thebibliography}{99}



\bibitem {hbn} 
    M. Engler, C. Lesniak, R. Damasch, B. Ruisinger, J. Eichler,  CFI {\bf 84}, 12 (2007)  
\bibitem {mos3d}
   A. V. Kolobov, J. Tominaga, {\it Two-Dimensional Transition-Metal Dichalcogenides}, Springer series in Materials Science (2016)
\bibitem{lee} 
    P. A. Lee, {\it Physics and chemistry of the materials with layered structure}, (Reidel, Dordrecht, 1976)
\bibitem{mehanika}   
    A. Falin {\it et. al.}, Nature Commun. {\bf 8},  (2017)  
 \bibitem{guo1} 
    G. Y. Guo, J. C. Lin, Phys. Rev. B {\bf 71}, 165402 (2005)
 \bibitem{wirt} 
    L. Wirtz, A. Marini, A. Rubio, Phys. Rev. Lett. {\bf 96}, 126104 (2006)
\bibitem{Thygesen} 
     J. Yan, K. W. Jacobsen, K. S. Thygesen, Phys. Rev. B {\bf 86}, 045208 (2012)
\bibitem{LouieMoS2} 
     D. Y. Qiu, F. H. da Jornada,  S. G. Louie, Phys. Rev. Lett. {\bf 111}, 216805 (2013)
\bibitem{koskelo} 
     J. Koskelo, G. Fugallo, M. Hakala, M. Gatti, F. Sottile,  P. Cudazzo, Phys. Rev. B {\bf 95}, 035125 (2017)
\bibitem{wannier} 
     G. H. Wannier, Phys. Rev.  {\bf 52}, 191 (1937) 
\bibitem{cuo2}   
     T. Kazimierczuk, D. Fr\"{o}hlich, S. Scheel, H. Stolz, M. Bayer, Nature {\bf 514}, 343 (2014)   
\bibitem{chernikov} 
     A. Chernikov, T. C. Berkelbach, H. M. Hill, A. Rigosi, Y. Li, O. B. Aslan, Phys. Rev. Lett. {\bf 113} 076802 (2014)     
\bibitem {exi-naj} %''Probing excitonic dark states in single-layer tungsten disulphide''
     Z. Ye, T. Cao, K. O'Brien, H. Zhu, Y. Wang, X. Yin, S. Louie,  X. Zhang,  Nature {\bf 513}, 214 (2014)    
\bibitem{kohn} 
     W. Kohn, J. M. Luttinger, Phys. Rev. {\bf 108}, 590 (1957)
\bibitem{wiser}
     N. Wiser, Phys. Rev.  {\bf 129}, 62 (1963)   
\bibitem{band-reno}  
     M. M. Ugeda  {\it et. al.}, Nature Materials {\bf 13}, 1091 (2014)  
\bibitem{ebk1} 
     M. Brack, R. Bhaduri, {\it Semiclassical Physics} (Addison-Wesley, 1977)
\bibitem{gapovi1} 
     A. Molina-Sanchez, D. Sangalli, K. Hummer, A. Marini, L. Wirtz, Phys. Rev. B {\bf 88}, 045412 (2013)
\bibitem{gapovi2} 
     H. P. Komsa, A. V. Krasheninnikov, Phys. Rev. B, {\bf 86}, 241201 (2012)
\bibitem{gapovi3} 
     D. Y. Qiu, F. H. Jornada,  S. G. Louie, Phys. Rev. Lett. {\bf 111}, 216805  (2013)
\bibitem{jeb} 
     T. Galvani, F. Paleari, H. P. C. Miranda, A. Molina-Sánchez, L. Wirtz, S. Latil, H. Amara, F. Ducastelle, Phys. Rev. B {\bf 94}, 125303 (2016)
\bibitem{QE} 
     P. Giannozzi, S. Baroni, N. Bonini, M. Calandra, R. Car, C. Cavazzoni, D. Ceresoli, G. L. Chiarotti, M. Cococcioni, I. Dabo, {\em et.al.}, J. Phys.: Conden. Matter {\bf 21}, 395502 (2009)
\bibitem{pseudopotentials} 
     N. Troullier and J. L. Martins, Phys. Rev. B {\bf 43}, 1993 (1991)
\bibitem{PZ} 
     J. P. Perdew and A. Zunger, Phys. Rev. B 23, 5048 (1981)
\bibitem{MPmesh} 
     H.J. Monkhorst and J.D. Pack, Phys. Rev. B {\bf 13}, 5188 (1976)
\bibitem{Hedin} 
     L. Hedin, Phys. Rev. {\bf 139}, 796 (1965)
\bibitem{Louie} 
     Mark S. Hybertsen, Steven G. Louie, Phys. Rev. B {\bf 34}, 5390 (1986)
\bibitem{indirekt} 
     G. Cassabois, P. Valin, B. Gil, Nature Photonics {\bf 10}, 262 (2016)
\bibitem{barisic} 
     S. Bari\v{s}i\'{c}, Phys. Rev. B  {\bf 5}, 932 (1972) 
\bibitem{walace}
     P. Wallace,  Phys. Rev.  {\bf 71}, 622 (1947)  
\bibitem{mahan} 
      G. D. Mahan, {\it Many-particle Physics} (Plenum Press, New York, 1990), 3rd ed.
\bibitem{nozier} 
      P. Nozieres and  D. Pines, {\it The Theory of Quantum Liquids I} (Addison-Wesley, New York, 1989).
\bibitem{kup} 
      I. Kup\v{c}i\'c, G. Nik\v{s}i\'{c}, Z. Rukelj, D. Pelc,  Phys. Rev. B  {\bf 94}, 075434 (2016) 
\bibitem {atomske-tablice}
      Kramida, A., Ralchenko, Yu., Reader, J., and NIST ASD Team (2014). NIST Atomic Spectra Database (ver. 5.2), [Online]. Available: http://physics.nist.gov/asd [2017, May 2]. National Institute of Standards and Technology, Gaithersburg, MD.  
\bibitem {T-D}
     T. Sander, E. Maggio,  G. Kresse,  Phys. Rev. B  {\bf 92}, 045209 (2015)      
\bibitem {vodik}
     X. Yang, S. Guo, F Chan,  Phys. Rev. A  {\bf 43}, 1186 (1991)
\bibitem{vito1} 
      V. Despoja, Z. Rukelj, L. Marusic, Phys. Rev. B {\bf 94}, 165446 (2016)
\bibitem {keldysh}
     L. Keldysh,  Pis'ma Zh. Eksp. Teor. Fiz.  {\bf 29}, 716  (1979) 
\bibitem{cudazzo} 
      P. Cudazzo, I. V. Tokatly, A. Rubio, Phys. Rev. B {\bf 84}, 085406 (2011)
\bibitem{vignale} 
       G. Giuliani and  G. Vignale, {\it Quantum Theory of the Electron Liquid } (Cambridge, New York, 2008).
\bibitem{gap-broj} 
       L. Schue {\it et. al.}, Nanoscale {\bf 8}, 6986 (2016)
\bibitem{piret} 
        A. Pierret {\it et. al.}, Phys. Rev. B {\bf 89}, 035414 (2014)
\bibitem{vito2}  
        Z. Rukelj, A. Strkalj, V. Despoja, Phys. Rev. B {\bf 94}, 115428 (2016)
\bibitem{wkb} 
        S. H. Dong, {\it Wave Equations in Higher Dimensions} (Springer, 2011)
\bibitem{ln-pot-sch} 
       K. Eveker, D. Grow, B. Jost, C. E. Monfort, K. W. Nelson, C. Stroh, R. C. Witt, Am. J. Phys. {\bf 58}, 1183 (1990)
\bibitem{prl1} 
 T. Olsen, S. Latini, F. Rasmussen, K. S. Thygesen, Phys. Rev. Lett. {\bf 116} 056401 (2016)       
 \bibitem{mote-exc} 
       I. G. Lezama, A. Arora, A.  Ubaldini, C. Barreteau, E. Giannini, M.  Potemski,  A. F. Morpurgo, Nano Lett., {\bf 4} 2336 (2015) 
\bibitem{mos-exc} 
       K. F. Mak, K. He, C. Lee, G. H. Lee, J. Hone, T. F. Heinz, J. Shan, Nature Materials, {\bf 12}, 207, (2013)
\bibitem{mos-exc2} 
       Y. Li, A. Chernikov, X. Zhang, A. Rigosi, H. M. Hill, A. M.  Zande, D. A. Chenet, E.-M. Shih, J. Hone,  T. F. Heinz, Phys. Rev. B {\bf 90}, 205422  (2014)       
   
   
   
   
 % \bibitem{ln-pot} I. R. Lapidus, Am. J. Phys. {\bf 49} 807 (1981)
  
  %onaj prvi prl s fitovima...

  

   

 
 % Exciton binding energy and nonrydberg series in monolayer mos

  



 
  
  
  

      
         
  % binding energy of trions and biexcitons in 2d izolatorima iz montekalro simulacija
   %\bibitem{rapid} M. Szyniszewski, E. Mostaani, N. D. Drummond, V. I. Fal'ko, Pys. Rev. B {\bf 95} 081301 (2017) 
  







%knjige







\end{thebibliography}
\end{document}